\documentclass[12pt]{iopart}
\usepackage{epsfig}    
\usepackage{setstack} 
\usepackage{inputenc}   
\usepackage{subfigure}  
\usepackage{amssymb}  
\usepackage{iopams}    
\begin{document}

\title{Effect of Pt impurities on the magnetocrystalline anisotropy of hcp Co: a first-principles study}
\author{C.J.~Aas$^1$, K.~Palot\'as$^2$, L.~Szunyogh$^{2,3}$, R.W.~Chantrell$^1$}
\address{$^1$ Department of Physics, University of York, York YO10 5DD, United Kingdom}
\address{$^2$ Department of Theoretical Physics, Budapest University of Technology and Economics, Budafoki \'ut 8.~H1111 Budapest, Hungary}
\address{$^3$ Condensed Matter Research Group of Hungarian Academy of Sciences, Budapest University of Technology and Economics, Budafoki \'ut 8.,~H-1111 Budapest, Hungary }


\begin{abstract}
In terms of the fully relativistic screened Korringa-Kohn-Rostoker method we investigate the variation in the magnetocrystalline anisotropy energy (MAE) of hexagonal close-packed cobalt with the addition of platinum impurities. In particular, we perform calculations on a bulk cobalt system in which one of the atomic layers contains a fractional, substitutional platinum impurity. Our calculations show that at small concentrations of platinum the MAE is reduced, while at larger concentrations the MAE is enhanced. This change of the MAE can be attributed to an interplay between on-site Pt MAE contributions and induced MAE contributions on the Co sites. The latter ones are subject to pronounced, long-ranged Friedel-oscillations that can lead to significant size effects in the experimental determination of the MAE of nano-sized samples.
\end{abstract}

\maketitle

\section{Introduction}
Cobalt alloys, such as CoPt or CoPd, are ubiquitous in the field of magnetic recording and of particular interest to the field of ultrafast magneto-optics \cite{lu}. In terms of magnetic recording, increasing areal densities require decreased grain size, which in turn requires increasing values of magnetocrystalline anisotropy energy (MAE) to ensure thermal stability of written information \cite{weller}. Currently this is achieved using CoPt alloys with perpendicular anisotropy. Consequently an understanding of the origin of the MAE in CoPt is an important practical problem. Since the magnetic properties of these alloys are highly sensitive to the amount and the spatial distribution of the Pt content, understanding the effects of alloying is an important issue.  The effects on the magnetic properties of CoPt as functions of the platinum content have been studied extensively, both theoretically~\cite{staunton} and experimentally \cite{klokholm, coexp}.  Moreover, in recent experimental work~\cite{jaworo} it was demonstrated that the magnetocrystalline anisotropy energy (MAE) of cobalt can be tuned by letting platinum impurities migrate into the cobalt system.  Generally it is agreed that the addition of platinum to a magnetic material, such as Fe or Co, influences the magnetic properties, in particular, the MAE of the material primarily through the strong spin-orbit coupling of Pt \cite{SOCCoPt}.\\*
 
The aim of the present work is to elucidate from first principles the effect on the MAE of bulk hcp Co by the addition of platinum. To this end, we use the fully relativistic screened Korringa-Kohn-Rostoker (SKKR) method as combined with the coherent-potential approximation (CPA), which is well suited to describing substitutional alloys \cite{kkrcpa}. Our model focuses on Pt alloying in a (0001) atomic plane of a hcp Co bulk system, from the case of an impurity to the case of a complete filling of the layer by Pt. After briefly discussing the computational methods we present the calculated MAE as a function of the Pt concentration and analyze the results
in terms of layer- and species-resolved contributions to the MAE. We note that recording media are complex alloy systems, often containing Cr to promote grain boundary separation. It is often found that the maximum MAE as a function of Pt concentration is limited by, for example, the formation of new phases \cite{toney} or the presence of stacking faults \cite{SF1,SF2,SF3}.  Here we are concerned only with the intrinsic enhancement of the MAE introduced by the Pt impurities. Remarkably, this analysis highlights the role of long-ranged Friedel oscillations in forming the MAE of the system. Specifically, we demonstrate a layer dependence of the valence charge which makes the effect of the Pt impurities long-ranged. This might have significant impact on the determination of the MAE of thin film samples corresponding to the systems studied in this work. In particular it might be expected to give rise to finite size effects in the MAE of granular thin films for magnetic recording which would become more significant as the grain size is reduced.

\section{Computational details}

The central feature of the SKKR method is the evaluation of the electronic Green's function of a layered system. Here, a layered system refers to a system exhibiting two-dimensional translational symmetry in each (infinite) atomic plane, but in which there are no symmetry requirements along the third axis. From the Green's function one can then determine a number of physical quantities of interest, such as site-projected charges, spin- and orbital moments and  the total energy of the system. As the method is well documented elsewhere \cite{SKKR,KKR-Hubert}, here we present only some details of our calculations. The calculations were performed within the local spin-density approximation (LSDA) of density-functional theory (DFT) as parametrised by Vosko {\em et al.}~\cite{voskoCJP80} The effective potentials and fields were treated in the framework of the atomic sphere approximation (ASA). The substitutional Pt alloying was treated within the coherent potential approximation (CPA) \cite{kkrcpa,soven}.  As the LSDA fails in predicting the orbital moment and the MAE for hcp Co correctly, we employed a heuristic extension of the relativistic electron theory by the orbital polarisation (OP) correction \cite{brooks,eschrig, eschrig2}, as implemented within the KKR method by Ebert and Battocletti \cite{OP-Hubert}. The corresponding Kohn-Sham-Dirac equations were solved using a spherical wave expansion up to an angular momentum number of $\ell=3$, although it should be noted that the OP correction was applied only for the $\ell=2$ orbitals.\\*

The magnetocrystalline anisotropy energy was evaluated within the magnetic force theorem \cite{Jansen99}, in which the total energy of the system can be replaced by the single-particle (band) energy. Moreover, we employed the torque method \cite{MAEtorque1}, making use of the fact that, for a uniaxial system, the MAE, $K$, can be calculated up to second order in spin-orbit coupling as
\begin{equation}
 K = E(\theta=90^{\circ}) - E(\theta=0^{\circ}) = \left.\frac{dE}{d\theta}\right|_{\theta=45^{\circ}} \, ,
\label{eq:torque}
\end{equation}
where, in the case of hcp geometry, $\theta$ denotes the angle of the spin-polarisation with respect to the $(0001)$ direction, i.e., the direction perpendicular to the hexagonal planes. Note that the $\mathbf{\hat{z}}$-axis of the (global) frame of reference in our calculations is defined to be parallel to the $(0001)$ direction. Within the KKR formalism, $K$ can be decomposed into site- and species-resolved contributions, 
\begin{equation}
K = \sum_{i, \alpha} s_i^{\alpha} D^{(\alpha)}_i \; ,
\label{eq:Kdecomp}
\end{equation}
where $s_i^{\alpha}$ denotes the concentration of species $\alpha$ at site $i$ and $D^{(\alpha)}_i$ denotes the corresponding derivative of the band energy.
Using Lloyd's formula \cite{Lloyd67}, $D^{(\alpha)}_i$ can be calculated as~\cite{MAEtorque2}
\begin{equation}
D_i^{(\alpha)} = -\frac{1}{\pi} \mathrm{Im} \int^{\epsilon_F^{(\hat{n})}} d\epsilon  \: \mathrm{Tr} \left( \frac{\partial  \underline{t}_{i}^{(\alpha,\mathbf{\hat{n}})}(\epsilon)^{-1}}{\partial \theta} \: \underline{\tau}^{(\alpha,\mathbf{\hat{n}})}_{ii}(\epsilon) \right) \: ,
\label{eq:Dia}
\end{equation}
where $\epsilon_F^{(\hat{n})}$ is the Fermi energy and, in case species $\alpha$ occupies site $i$,  $\underline{t}_i^{(\alpha,\mathbf{\hat{n}})}(\epsilon)$ and $\underline{\tau}^{(\alpha,\mathbf{\hat{n}})}_{ii}(\epsilon)$ stand for the angular momentum matrices of the single-site $t$ operator and the site-diagonal scattering path operator, respectively. All these quantities are calculated at the direction of the magnetisation $\mathbf{\hat{n}}=\left(  \frac{1}{\sqrt{2}}, 0, \frac{1}{\sqrt{2}} \right)$, corresponding to $\theta=45^{\circ}$ in Eq.~(\ref{eq:torque}). The derivative of the $t$-matrix is evaluated as described in~\cite{spinwaves}. The energy integral in Eq.~(\ref{eq:Dia}) can be accurately performed by sampling 20 energy points on an asymmetric mesh along a semi-circle contour in the upper complex semi-plane. In order to achieve an accuracy within 5 \% for the MAE, a sufficiently dense mesh in the two-dimensional Brillouin zone (2D-BZ) was used to evaluate $\underline{\tau}^{(\alpha,\mathbf{\hat{n}})}_{ii}(\epsilon)$: at the energy point closest to the Fermi energy, we used 5764 $k$-points in the irreducible wedge of the 2D-BZ, corresponding to more than 34~000 $k$-points in the full 2D-BZ. Due to the two-dimensional translational symmetry of the system, the MAE should be related to a 2D unit cell, therefore, in the following the index $i$ in Eq.~(\ref{eq:Kdecomp}) is used to label atomic layers.\\* 

\begin{figure}[htb]
\begin{center}
\includegraphics[scale=0.3]{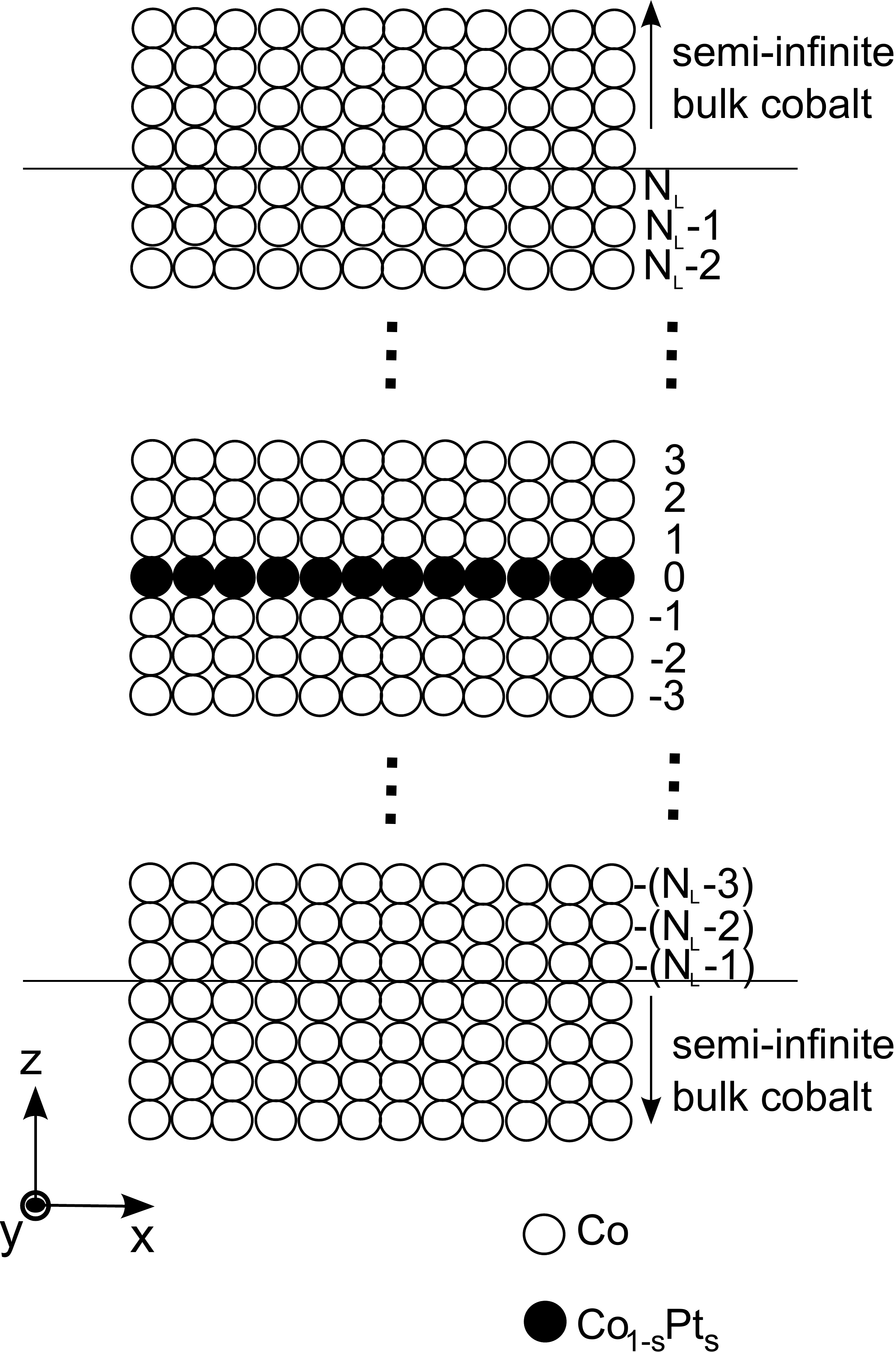}
\caption{ \small Sketch of the geometry of the system containing $N_L$ hcp unit cells, i.e., $2N_L$ atomic layers wedged between
two perfect semi-infinite bulk Co systems. In the zero-indexed layer (black circles) a random substitutional alloy with platinum, Co$_{1-s}$Pt$_s$, is considered.
Note that the $\mathbf{\hat{z}}$-axis is defined to be parallel to the $(0001)$ direction of the hcp crystal. \normalsize\label{coptsystem}}
\end{center}
\end{figure}

The SKKR method as applied to layered systems requires the system to be divided into a middle region wedged between two semi-infinite bulk regions. Adhering to this requirement, the effect of platinum alloying in a single atomic layer of bulk hcp Co was investigated by considering a layered system as shown in Fig.~\ref{coptsystem}.  Each atomic layer in the semi-infinite bulk regions corresponds to pure hcp Co bulk. Since the middle region needs to contain  an integer number of unit cells and since each unit cell spans two atomic layers, this region consists of $2N_L$ hexagonal Co layers stacked along the $(0001)$ direction. In one of the two central layers of the middle region, namely, in the one indexed by 0 in Fig.~\ref{coptsystem}, a fraction $s$ of the Co atoms are replaced by Pt atoms. From here on, this layer will be referred to as the \emph{impurity layer}. It should also be mentioned that in this  work no attempts are made to trace any structural relaxation effects of the hcp Co lattice caused by Pt impurities.\\*

To take into account relaxation of the effective potentials and fields, we performed self-consistent calculations with $N_L=14$, i.e.~for 28 layers in total. One important consequence of the geometrical construction shown in Fig.~\ref{coptsystem} is that the calculation of $K$ in Eq.~(\ref{eq:Kdecomp}) is confined to layers within the middle region, i.e., for $-(N_L-1) \le i \le N_L$. This means that the long-ranged Friedel oscillations that arise due to the presence of Pt impurities are necessarily truncated. In order to safeguard against any numerical artefacts caused by this truncation, we increased the number of atomic layers in the middle region until the layer-resolved MAE converged to within about 1 \% accuracy to the bulk Co MAE at the outer edges of the middle region. According to our calculations (see below), this condition requires $N_L = 40$, i.e., 80 atomic layers in total. We performed these calculations of the MAE by appending the perfect bulk potential of hcp Co to the layers $-39 \le i \le -14$ and $15 \le i \le 40$, i.e., neglecting self-consistency effects for these atomic layers. To check the accuracy of this approach, we compared $D_i^{(\mathrm{Co})}$ for atomic layer no. 14 (with relaxed self-consistent potential) with that for atomic layer no. $-14$ (with appended Co bulk potential) and obtained that the two values agree to within $0.02$ \%.

\section{Results and Discussion}
To test our computational method, we first determined the MAE of bulk hcp Co. Excluding the OP correction we obtained an easy-plane magnetisation and a MAE of 6.7 $\mu$eV/Co~atom, while including the OP correction we instead obtained an easy axis perpendicular to the hexagonal Co planes and a MAE of 84.4 $\mu$eV/Co. The latter result is in good agreement with the experimental value of 65.5 $\mu$eV \cite{stearns} and with the experimental easy axis being along the (0001) direction. Our result also compares well with that of Trygg {\em et al.} \cite{trygg95}, who calculated $K=110$~$\mu$eV for hcp Co using a full-potential LMTO method including OP correction.\\*

\begin{figure}[htp]
\begin{center}
\includegraphics[scale=0.35]{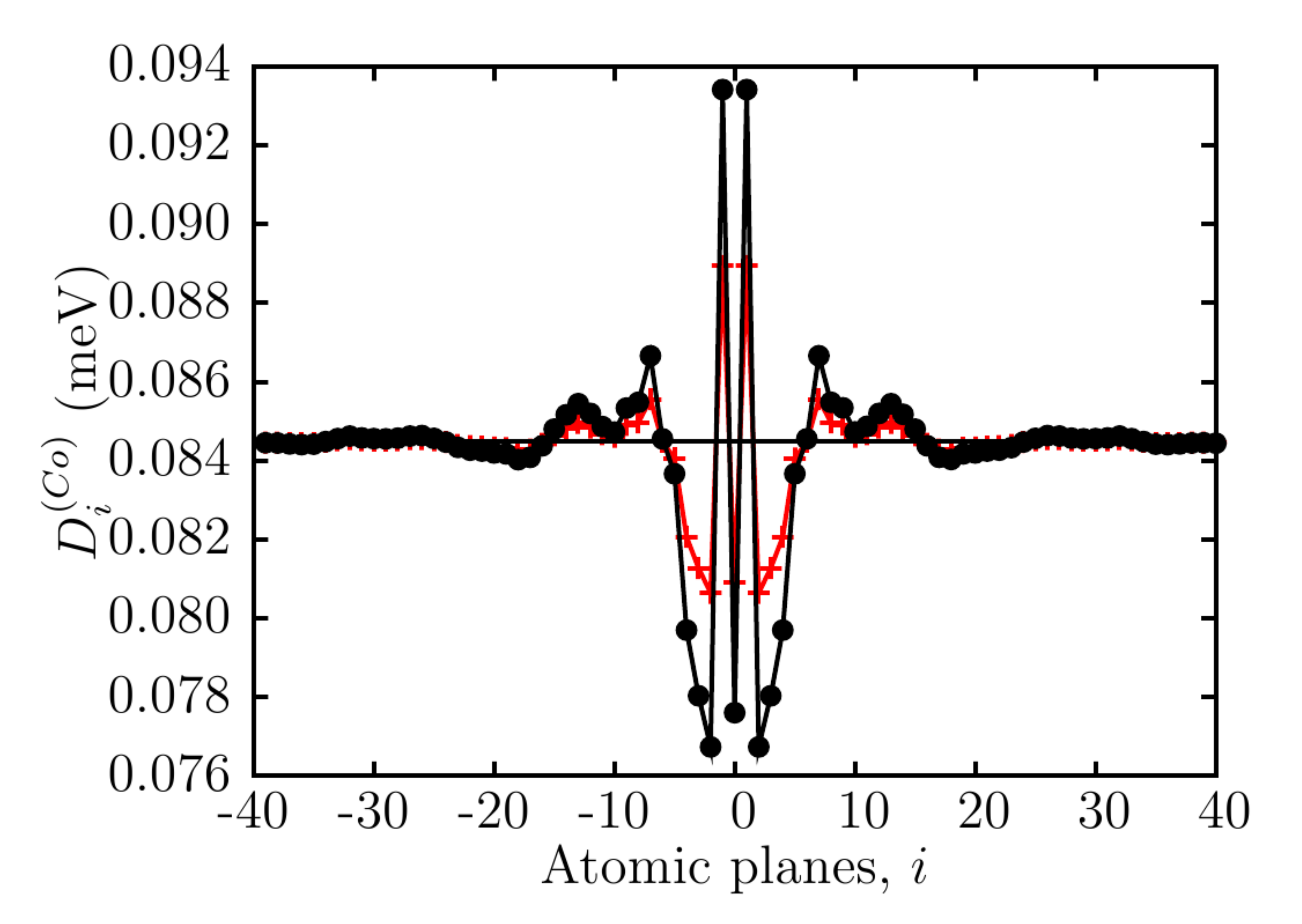}
\caption{
(Color online) Calculated layer-resolved Co contributions, $D_i^{(Co)}$, to the MAE, see Eq.~(\ref{eq:Dia}),
across the system shown in Fig.~\ref{coptsystem} for $s=0.01$ (red $+$) and $s=0.02$ (black $\bullet$).
The MAE of bulk Co is indicated by the solid black horizontal line.
Solid lines connecting the symbols serve as guides for the eye.
\label{layerMAEfig}}
\end{center}
\end{figure}

As described in Section 2, we performed calculations of the MAE of a bulk Co system in which a single layer has been substitutionally alloyed by Pt in a fraction of $0 < s \le 1$. The layer-resolved Co contributions to the MAE, $D_i^{(Co)}$, see Eqs.~(\ref{eq:Kdecomp}) and (\ref{eq:Dia}), are shown in Fig.~\ref{layerMAEfig} for $s=0.01$ and $s=0.02$. Remarkably, even such small amounts of Pt induce large fluctuations in $D_i^{(Co)}$: in the impurity layer ($i=0$) and in the Co layers near the impurity layer ($1 \le \mid \! i \mid \le 4$) the relative changes of $D_i^{(Co)}$ with respect to the bulk Co MAE reach 10~\%.  In particular, the Co contribution from layers $i=\pm$1 is enhanced to nearly 94 $\mu$eV, while that those from layers $i=\pm$2 are reduced to nearly 76 $\mu$eV for $s=0.02$. For layers further away from the impurity layer ($\mid \!i \mid \ge 5$), oscillations in $D_i^{(Co)}$ with rapidly decreasing amplitude can be seen. Reassuringly, the layer-resolved Co contributions approach the bulk Co MAE towards the outer edges of the middle region chosen in our calculations ($i \rightarrow N_L=40$ and $i \rightarrow -(N_L-1)=-39$). The mirror symmetry around the impurity layer, $D_i^{(Co)} = D_{-i}^{(Co)}$, is also fulfilled with a high accuracy.\\*

\begin{figure}[!htp]
\begin{center}
\subfigure{\includegraphics[scale=0.35]{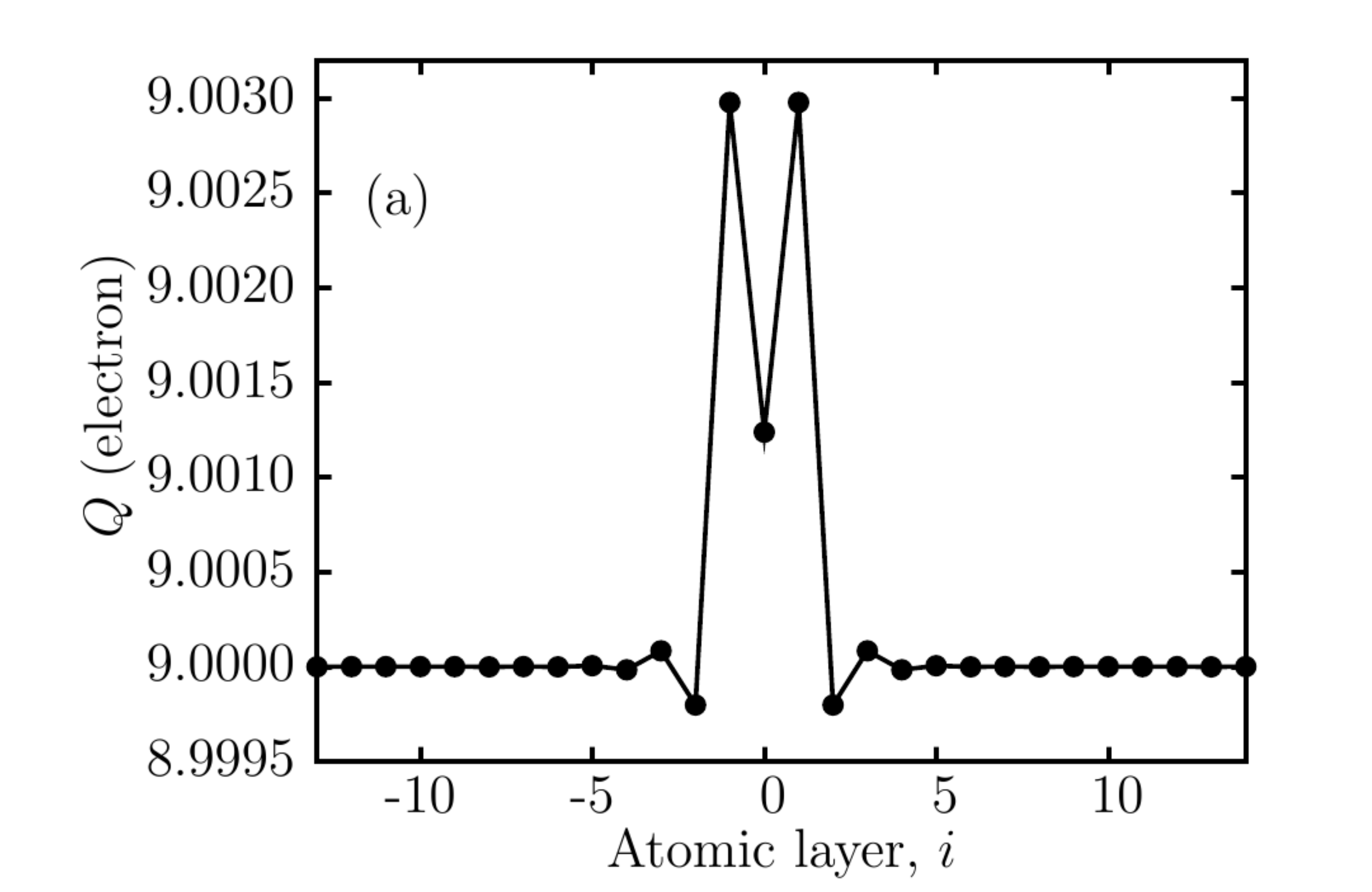}}
\subfigure{\includegraphics[scale=0.35]{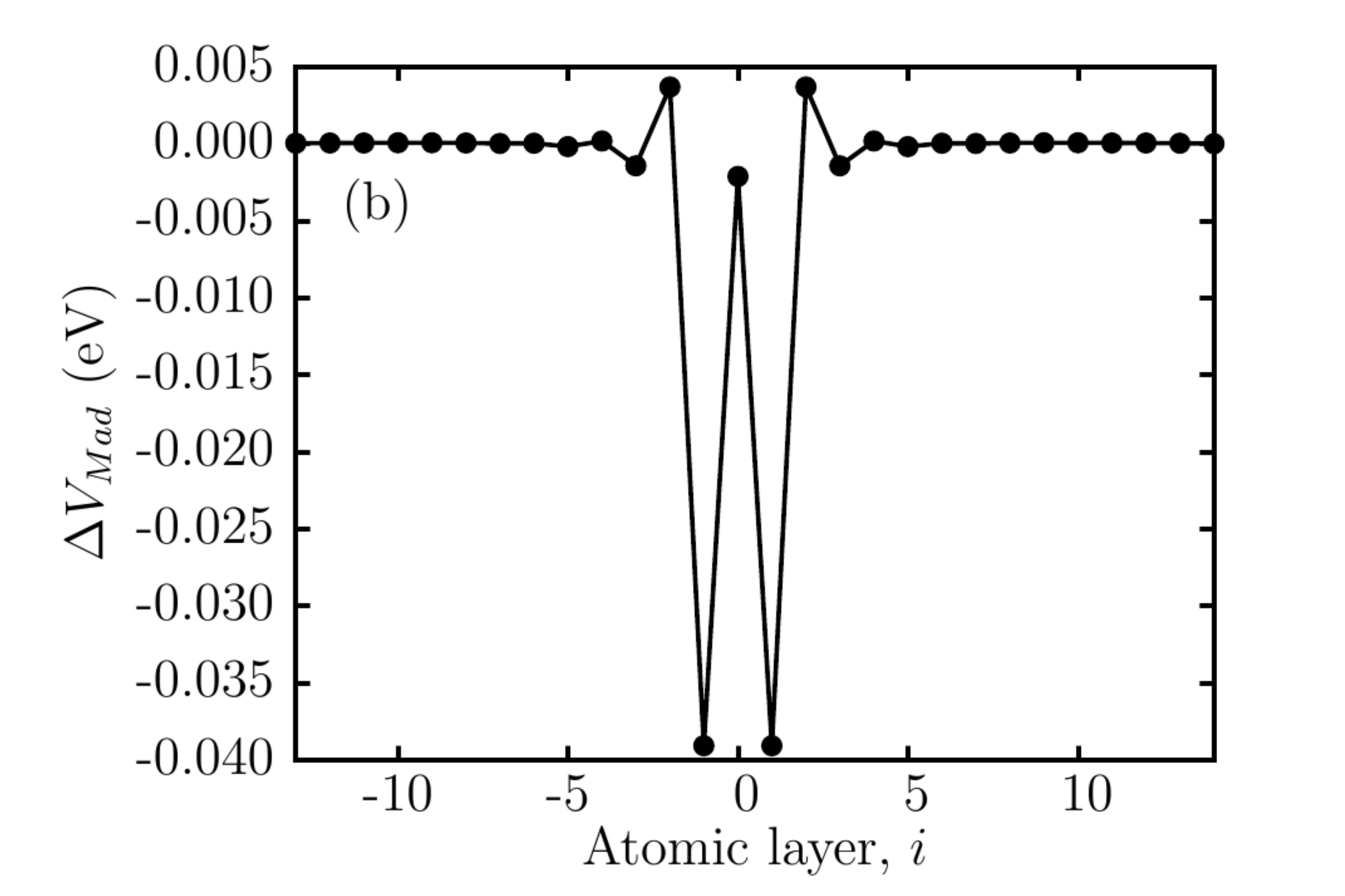}}
\caption{(a) Calculated valence charge on the Co atoms, $Q$, and (b) relative shift of the Madelung potentials
with respect to the bulk case, $\Delta V_{Mad}$, for layers $-13 \le i \le 13$  and for a Pt concentration, $s=0.02$.
Solid lines serve as guides for the eye.
\label{charge_and_madelung_layers}}
\end{center}
\end{figure}

Our earlier studies of the MAE of impurities~\cite{szunyogh98,szilva08}, justified that the MAE is extremely sensitive to the presence of Friedel oscillations in the charge density. It is, therefore, tempting to relate Fig.~\ref{layerMAEfig} to the change in the valence charge on the Co atoms with respect to the distance from the impurity layer. From Fig.~\ref{charge_and_madelung_layers}(a) we can see that the Co atoms in the impurity layer and, in particular, the Co atoms in layers adjacent the impurity layer gain some extra charge, while the charge transfer to more distant Co layers drops rapidly. The energy shift of the layer-resolved Co valence band position is well  described by the layer-resolved change in the Madelung potential. (Note that within the ASA the Madelung potential in each atomic plane is a constant.) As is obvious from Fig.~\ref{charge_and_madelung_layers}(b), an enhanced (reduced) charge at the Co sites is accompanied with a downward (upward) shift of the valence states. Comparing with Fig.~\ref{layerMAEfig}, this shift of the valence states correlates directly with the MAE contributions of the Co layers adjacent the impurity layer, but, clearly enough, the changes in the MAE contributions from more distant Co layers are also subject to fine details of the valence states influenced by the Pt alloying.\\*

\begin{figure}[!htb]
\begin{center}
\subfigure{\includegraphics[scale=0.35]{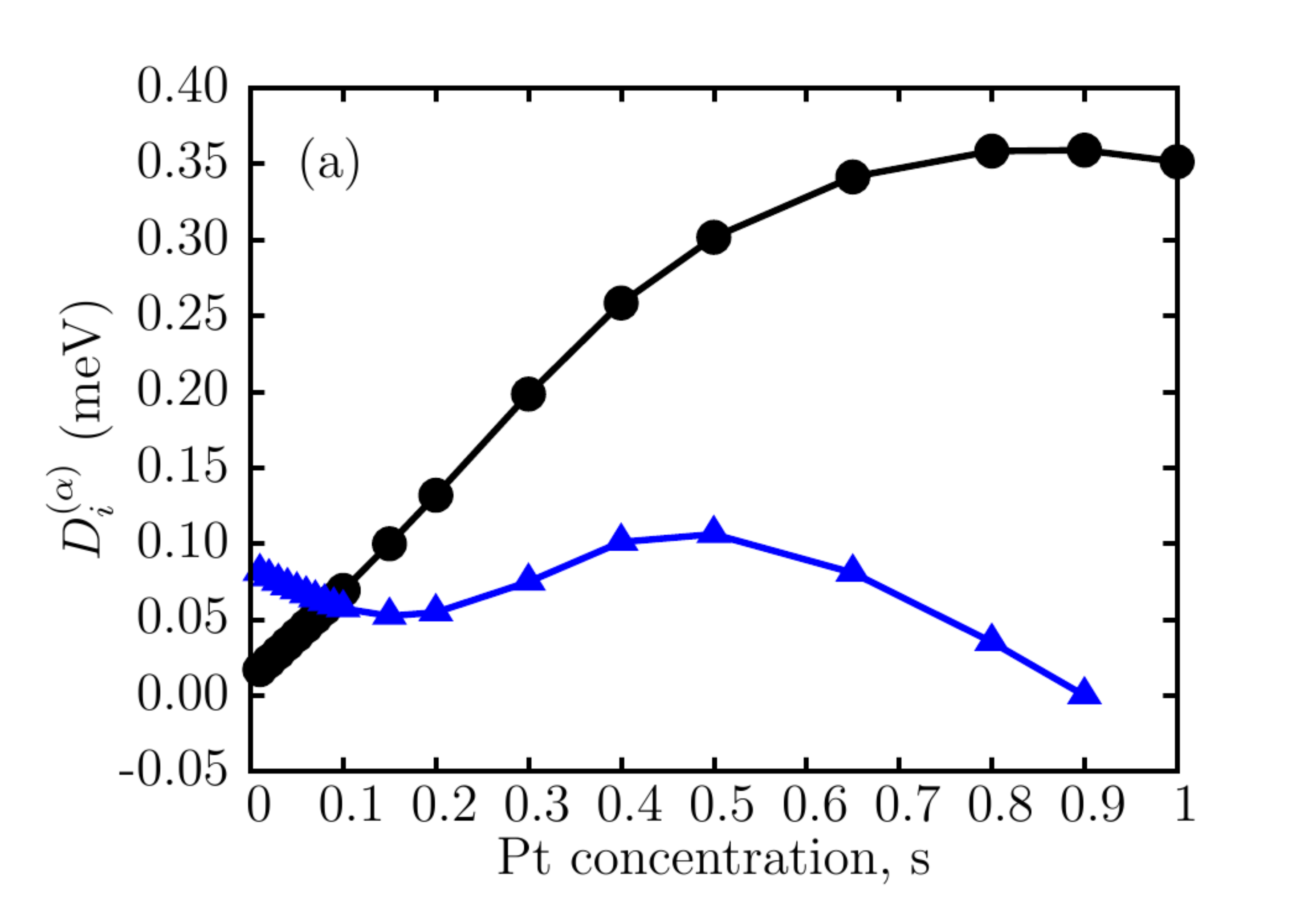}}
\subfigure{\includegraphics[scale=0.35]{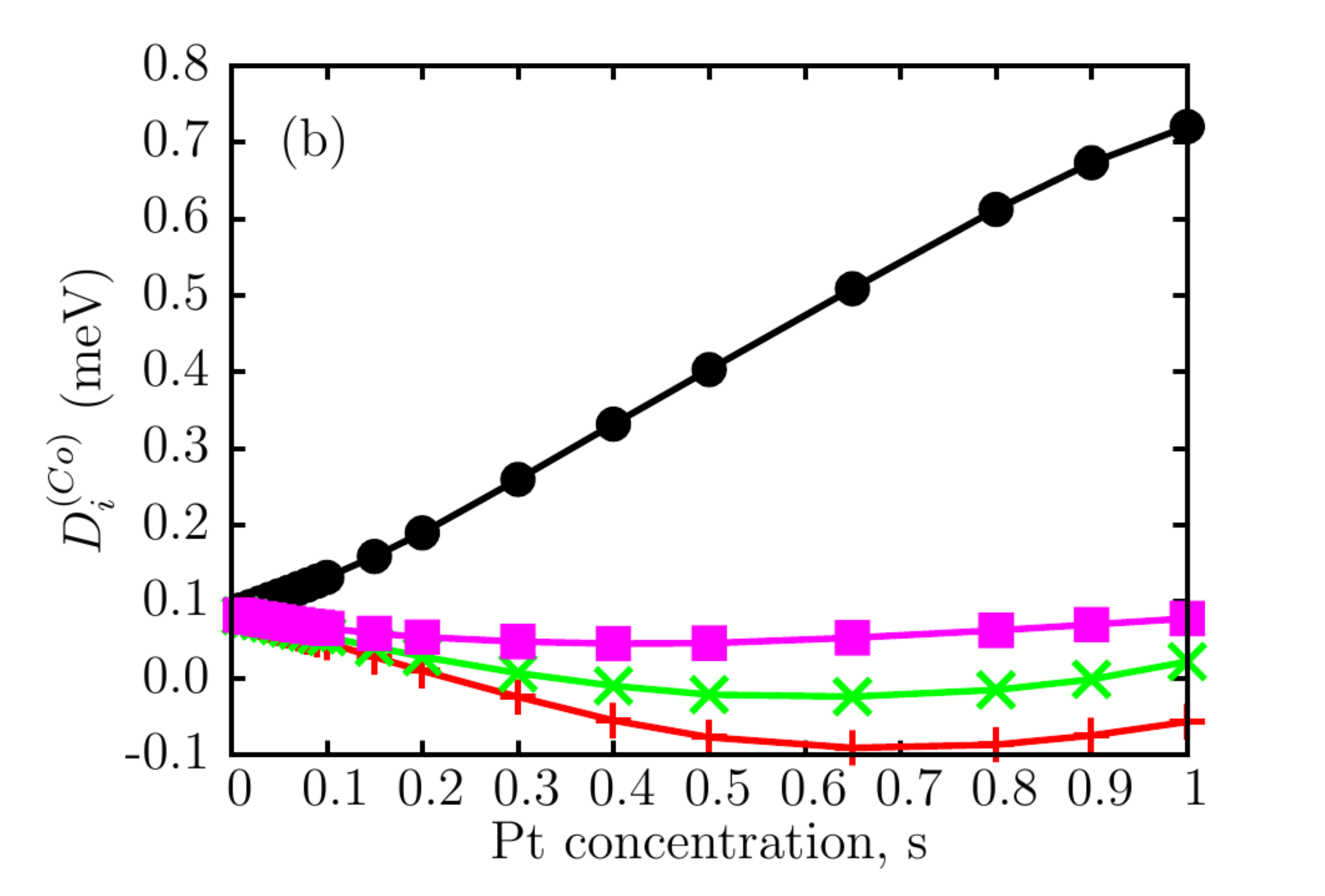}}
\caption{(Color online)
Calculated species-resolved MAE contributions (a) for the impurity layer: $D_0^{(Pt)}$ (black {\large $\bullet$}) and $D_0^{(Co)}$ (blue $\blacktriangle$) and (b) for Co layers: $D_1^{(Co)}$ (black {\large $\bullet$}), $D_2^{(Co)}$ (red +), $D_3^{(Co)}$ (green $\times$) and $D_4^{(Co)}$ (purple $\blacksquare$).
Solid lines connecting the symbols serve as guides for the eye.
\label{comparison4}}
\end{center}
\end{figure}

The effect on the species-resolved MAE by alloying with Pt is demonstrated for the whole range of $s$ in Fig.~\ref{comparison4}, showing $D_i^{(Co)}$ for layers $0\leq i \leq 4$, together with the direct contribution of Pt, $D_0^{(Pt)}$. In the impurity layer $i=0$, see Fig.~\ref{comparison4}(a), the Co contribution $D_0^{(Co)}$ is reduced by the addition of Pt for concentrations up to about $s=0.15$ and then enhanced for concentrations $0.15 < s < 0.50$. For concentrations $s > 0.50$, $D_0^{(Co)}$ is again reduced with increasing $s$ and at $s\approx 0.85$ $D_0^{(Co)}$ even becomes negative. Note that $D_0^{(Co)}$ for $s \rightarrow 1$ (not calculated here) would correspond to the contribution of a single Co atom in a pure Pt layer which, in general, would differ from zero. The on-site platinum contribution, $D_0^{(Pt)}$, approaches the very small value of 0.01~meV as $s\rightarrow0$, rapidly increases up to 0.30 meV at $s\approx0.5$ and then saturates at $D_0^{(Pt)} \approx 0.35$~meV for larger $s$.\\*

It can be inferred from Fig.~\ref{comparison4}(b), that Pt alloying most dramatically influences the Co contribution at the layers adjacent to the impurity layer: $D_1^{(Co)}$ increases almost linearly from the bulk MAE at $s=0$ to about 0.7~meV at $s=1$. As already seen in Fig.~\ref{layerMAEfig}, the Co contributions $D_i^{(Co)}$ from layers further out ($2 \le i \le 4$) decrease with increasing $s$ and even becomes negative at $s \approx 0.25$ for $i=2$ and 3. For $s > 0.5$ these contributions show a modest increase, but $D_2^{(Co)}$ still remains negative.\\*

\begin{figure}[!htp]
\begin{center}
\subfigure{\includegraphics[scale=0.35]{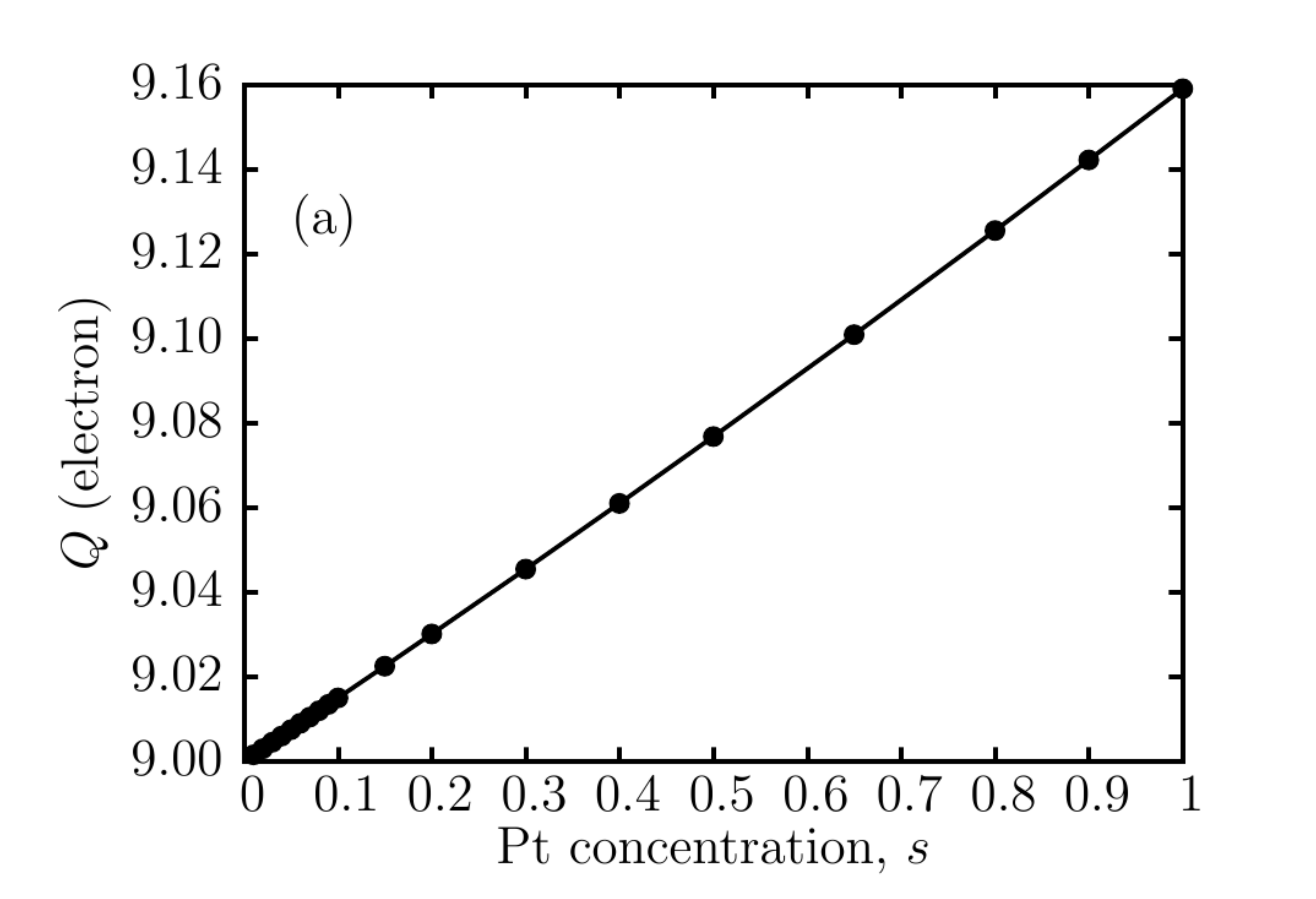}}
\subfigure{\includegraphics[scale=0.35]{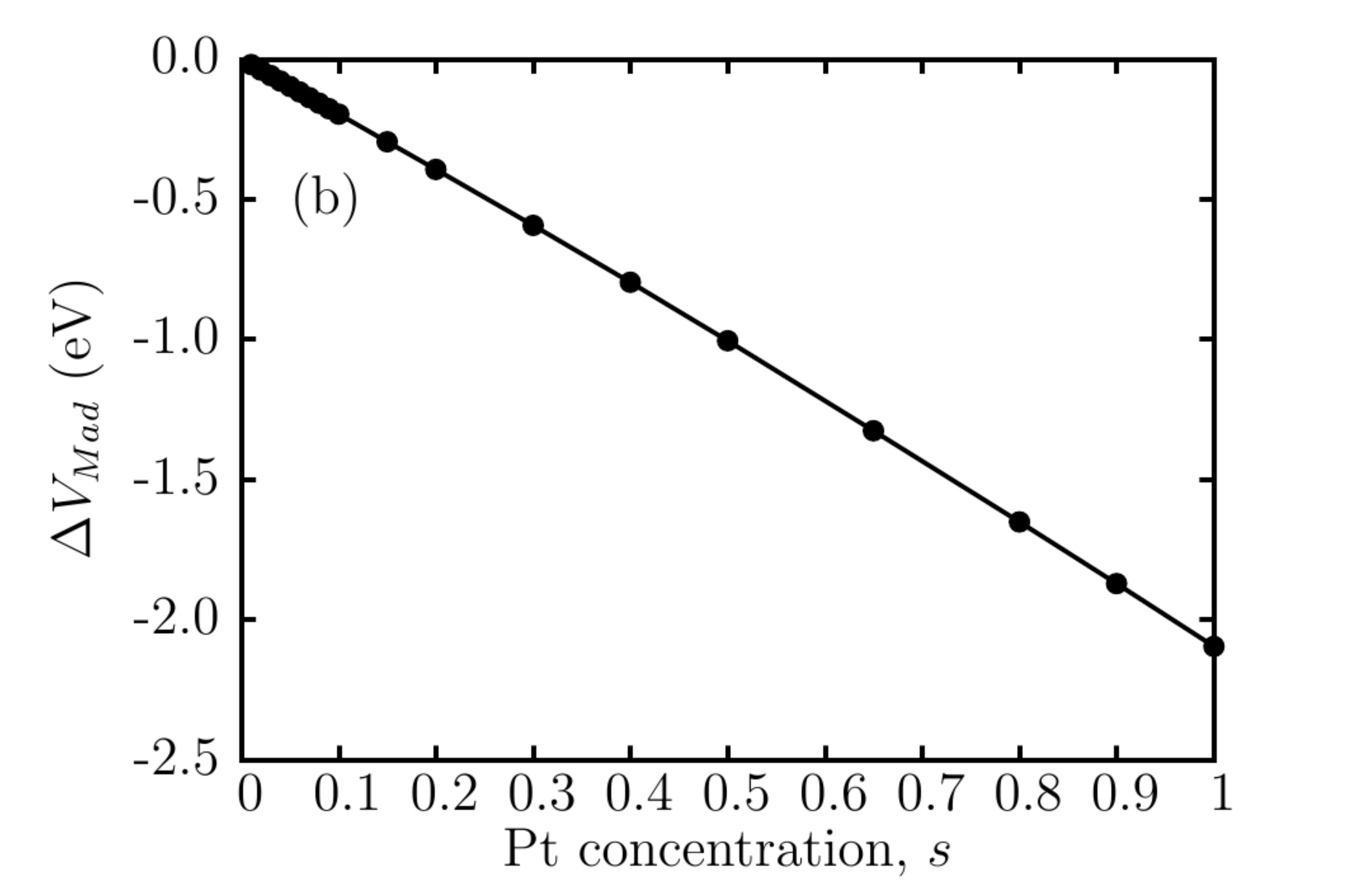}}
\caption{(a) Calculated valence charge on the Co atom, $Q$,  and (b) relative shift of the Madelung potential
with respect to the bulk case, $\Delta V_{Mad}$, for layer 1 as a function of the Pt concentration, $s$.
Solid lines serve as guides for the eye.
\label{charge_and_madelung_layer1}}
\end{center}
\end{figure}

It is worth investigating the change of valence states projected onto the Co atoms in layer 1. From Fig.~\ref{charge_and_madelung_layer1}(a) it is obvious that the valence charge at this Co atom increases almost linearly with $s$ from 9.00 $e$ to 9.16 $e$. This increase in the valence charge is necessarily accompanied by a downshift of the corresponding valence states, as characterized by the change in the Madelung potential, which is also linear $s$, see  Fig.~\ref{charge_and_madelung_layer1}(b). The large enhancement of the MAE contribution from layer 1, $D_1^{(Co)}$, can therefore be related directly to the monotonic shift of the corresponding valence states.\\*

While the species- and layer-resolved contributions to the MAE are very illuminating for a microscopic description of the variations in the MAE, from an experimental point of view only the MAE of the whole system can be accessed. Here, this means considering the MAE of the entire middle region illustrated in Fig.~\ref{coptsystem} for $N_L=40$. In order to extract the change in this MAE induced by the Pt impurities, we define the excess MAE, $\Delta K(s)$, by subtracting the MAE of the 'unperturbed' cobalt bulk layers,
\begin{equation}
  \Delta K(s) =  sD_{0}^{(Pt)} + (1-s)D_{0}^{(Co)}  + 2\sum_{i=1}^{40}D_{i}^{(Co)} - 81 K_{Co} \:,
\label{deltaKeq}
\end{equation}
where $K_{Co}$ is the calculated MAE of hcp bulk Co (84.4 $\mu$eV).  Note that we have taken into account the off-centre positioning of the impurity layer by doubling the Co contributions $D_{i}^{(Co)}$ for $i\in [1,40]$,
thus, in total, a system of 81 layers is considered.

\begin{figure}[!htp]
\begin{center}
\includegraphics[scale=0.35]{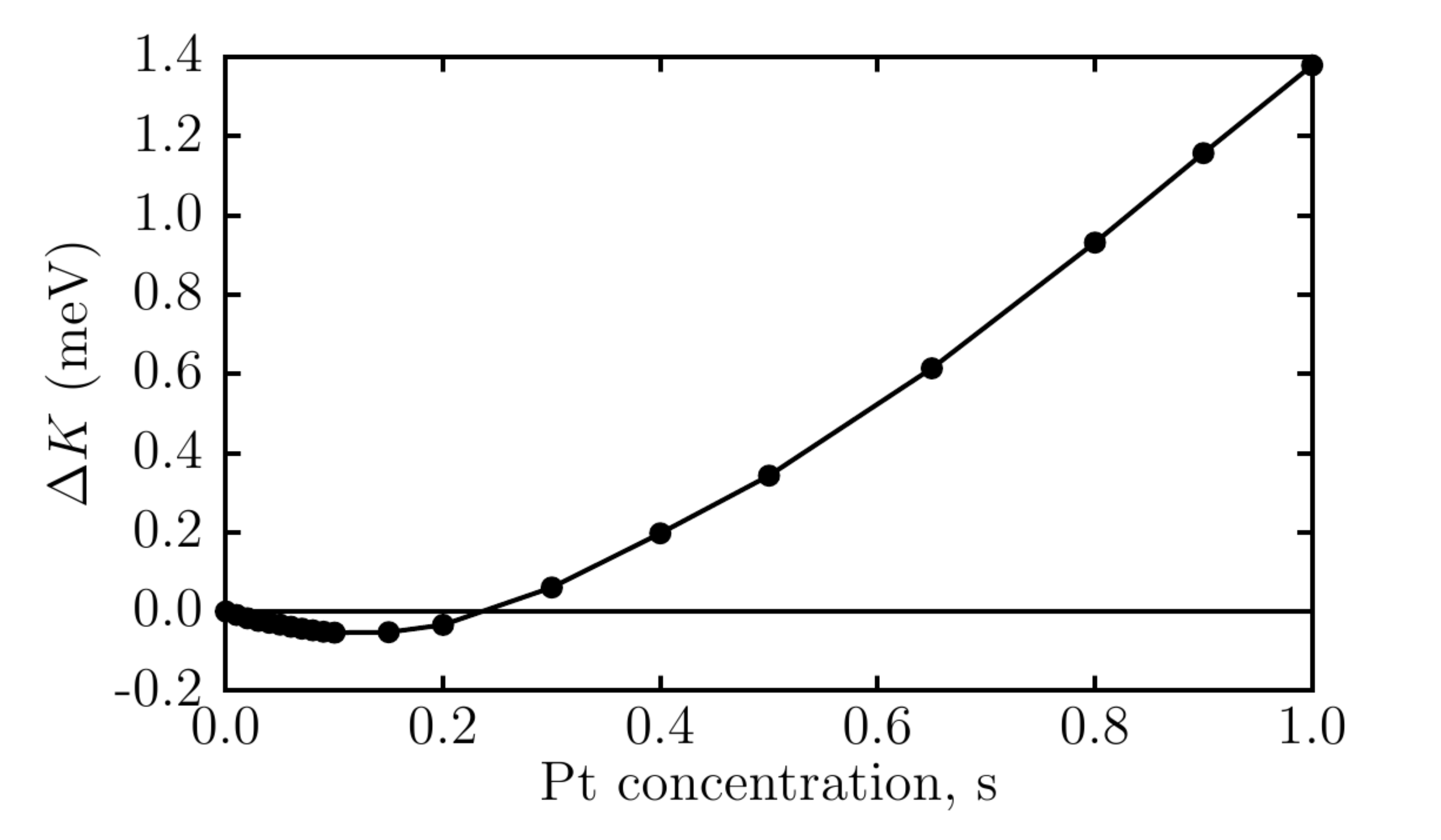}
\caption{The change in the MAE, $\Delta K$, of a system of 81 atomic layers, see Eq.~(\ref{deltaKeq}),
as a function of the Pt concentration, $s$. The solid line connecting the symbols serves as a guide for the eye.
\label{comparison2}}
\end{center}
\end{figure}

$\Delta K$ is shown as a function of $s$ in Fig.~\ref{comparison2}, demonstrating that for small concentrations of platinum ($s < 0.24$) the addition of platinum to bulk cobalt actually reduces the total MAE of the system by about 80~$\mu$eV.  This is in strong contrast to the on-site contribution of Pt, $D_{0}^{(Pt)}$, being positive for all values of $s$ as seen in Fig.~\ref{comparison4}(a). The reduction in MAE for low $s$, therefore, stems from the decrease in the cobalt contributions $D_{i}^{(Co)}$ with increasing $s$, in particular, for $i=0,2,3$ and $4$, see Fig.~\ref{comparison4}. $\Delta K$ becomes positive for $s>0.24$ as the increasing on-site contribution, $D_{0}^{(Pt)}$, gets larger weight (note that it is multiplied by $s$) and due to the large enhancement of $D_1^{(Co)}$. At $s=1$, $\Delta K = 1.4$ meV, which is approximately four times the on-site platinum contribution, $D_0^{(Pt)} \approx 0.35$ meV for $s=1$.\\*

\begin{figure}[htp]
\begin{center}
\includegraphics[scale=0.35]{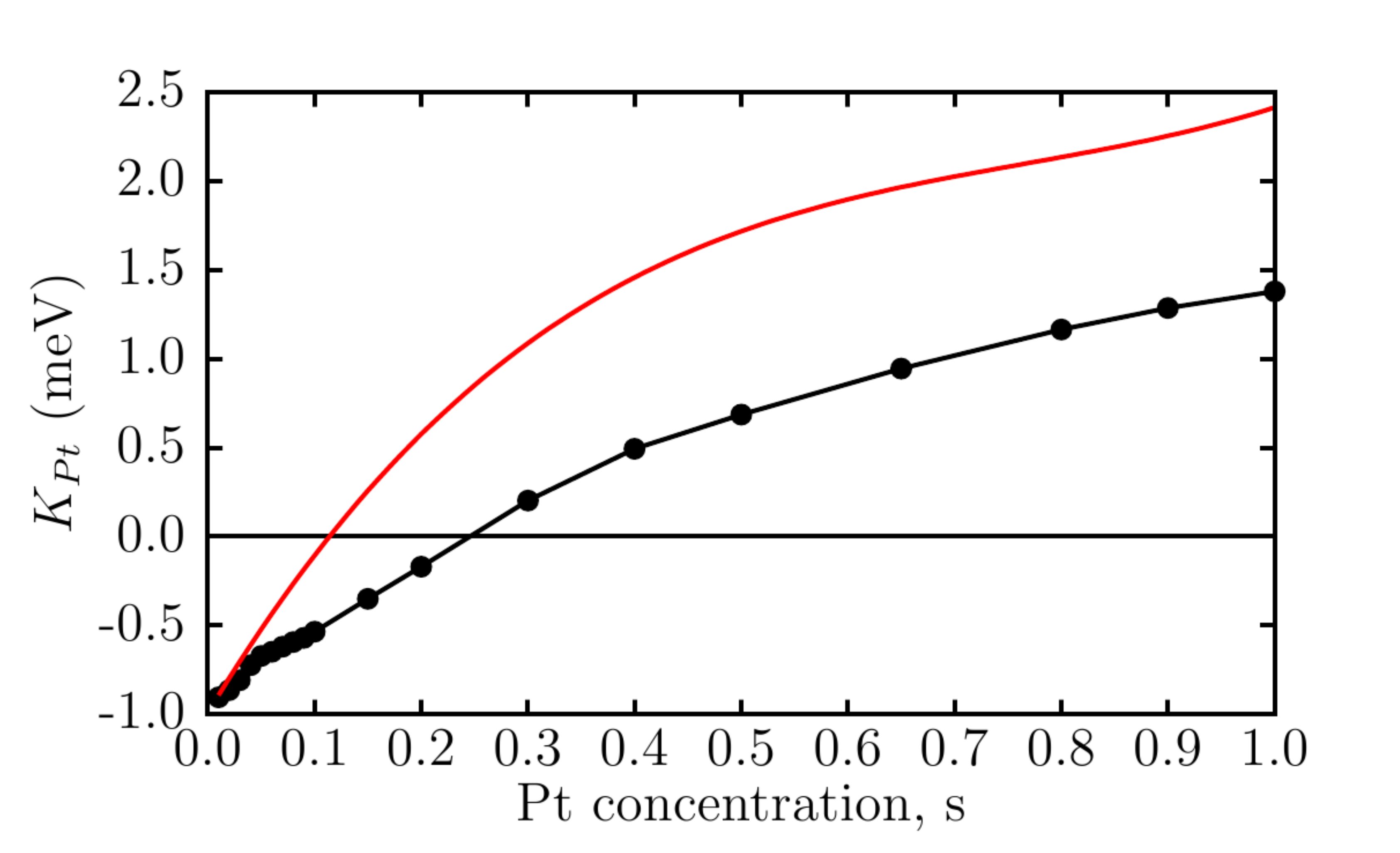}
\caption{(Color online) Black circles: calculated change in the MAE per Pt atom, $\overline{K}_{Pt}$, see Eq.~(\ref{KavPteq}),
as a function of the Pt concentration, $s$.
The black solid line connecting the symbols serves as a guide for the eye.
Red solid line: the change in the MAE per Pt atom added,  ${K}_{Pt}$, see Eq.~(\ref{KPteq}),
as calculated from a polynomial fit of $\Delta K(s)$ in Fig.~\ref{comparison2}.
 \normalsize\label{comparison3}}
\end{center}
\end{figure}

For nano-sized systems, it might be of interest to consider {\em the change in the MAE per platinum atom in the system}, $\overline{K}_{Pt}$, defined by
\begin{equation}
  \overline{K}_{Pt}(s) = \frac{\Delta K(s)}{s} \: ,
\label{KavPteq}
\end{equation}
and also, {\em the change in the MAE per platinum atom added to the system}, $K_{Pt}$, obviously given by
\begin{equation}
  K_{Pt}(s) = \frac{d \left( \Delta K(s) \right)}{ds} \: .
\label{KPteq}
\end{equation}
We obtained $K_{Pt}(s)$ by fitting a fourth-order polynomial to the function $\Delta K(s)$ in Fig.~\ref{comparison2} and then finding the derivative of this function analytically.  As apparent from Fig.~\ref{comparison3}, both $\overline{K}_{Pt}$ and ${K}_{Pt}$ are monotonically increasing with increasing $s$, starting with the same value of about $-1$ meV at $s=0$ (see later).  It follows directly from Fig.~\ref{comparison2}, that $\overline{K}_{Pt}$ crosses zero at $s \approx 0.24$, while $\overline{K}_{Pt}$ crosses zero at $s \approx 0.11$ (i.e.~where the function $\Delta K(s)$ reaches its minimum). For a complete platinum layer immersed in bulk cobalt, i.e.~for $s=1$, $\overline{K}_{Pt} = \Delta K \approx 1.4$ meV. A comparison with Fig.~\ref{comparison4} shows that about 25 \% of this value arises from the direct contribution of Pt, $D^{(Pt)}_1$, and the rest from the induced contributions at the Co atoms. Interestingly, the change in the MAE by addition of a Pt atom to the system, $K_{Pt}$, exhibits a surprisingly large value of about 2.5~meV at $s=1$. From Fig.~\ref{comparison4}(a) it can be inferred that the on-site Pt contribution $D_0^{(Pt)}$ has nearly zero slope in this region of $s$, thus, this large value of $K_{Pt}$ stems mainly from an increase in $D^{(Co)}_i$ for $1 \le \mid i \mid \le 4$ near $s=1$.\\*

In the limit $s\rightarrow 0$, corresponding to the case of a single Pt impurity in bulk Co, $\overline{K}_{Pt}$ and ${K}_{Pt}$ should be identical, since for small $s$ the function $\Delta K(s)$ exhibits, in principle, a linear dependence. This is fairly well confirmed by our calculations. $K_{Pt}(0)$ can then be expressed as
\begin{equation}
  K_{Pt}(0) =  D_{0}^{(Pt)}(0) - K_{Co}  +
  \sum_{i=-40}^{40} \left. \frac{dD_{i}^{(Co)}(s)}{ds} \right|_{s=0} \: .
\label{KPt0eq}
\end{equation}
The physical meaning of the above equation is that adding a Pt impurity to bulk Co has two effects on the MAE of the system: the first two terms, $D_{0}^{(Pt)}(0) - K_{Co}$, represent the direct contribution of a Co atom being replaced by a Pt atom, whereas the the last term of Eq.~(\ref{KPt0eq}) quantifies the induced change in the MAE contributions from the Co atoms that are not being replaced by Pt. Since the direct contribution is about -0.07~meV, see also Fig.~\ref{comparison4}(a), the value of $K_{Pt}(0)=-1$~meV can again only be explained by the induced Co contributions.\\*

\begin{figure}[htp]
\begin{center}
\includegraphics[scale=0.35]{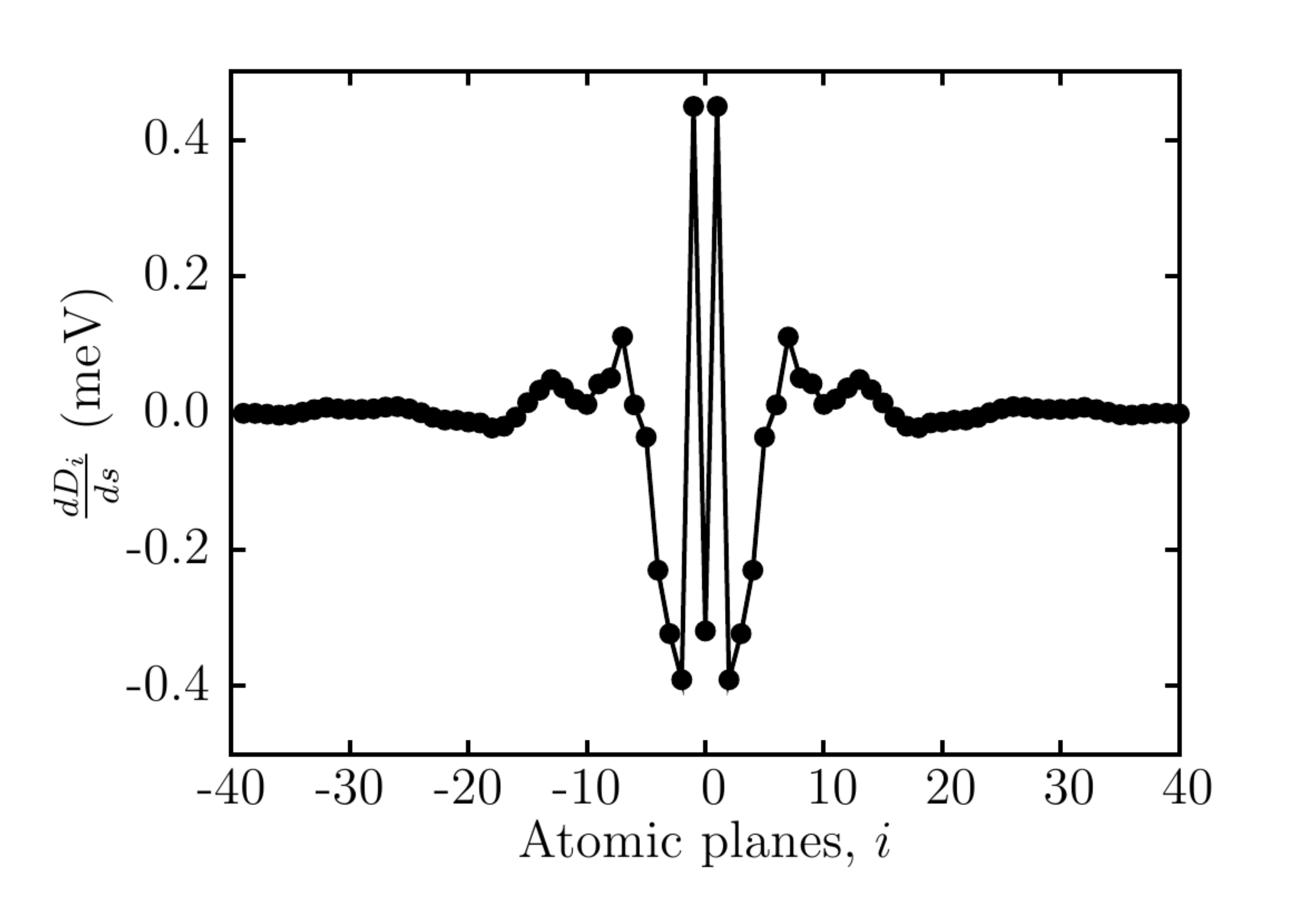}
\caption{\small Calculated derivatives of the layer-resolved Co contributions to the MAE, Eq.~(\ref{eq:Dder}),
 for the Pt concentration $s=0.02$. The solid line serves as a guide for the eye.
 \normalsize\label{layer_deriv_fig}}
\end{center}
\end{figure}

In  Fig.~\ref{layer_deriv_fig} we show the approximate layer-resolved derivatives calculated as
\begin{equation}
\left. \frac{dD_i^{(Co)}(s)}{ds} \right|_{s_j} = \frac{D^{(Co)}_i(s_{j+1})-D^{(Co)}_i(s_{j-1})}{s_{j+1}-s_{j-1}}
\; ,
\label{eq:Dder}
\end{equation}
for $s_j=0.02$ with $j$ indexing the Pt concentrations in ascending order. Clearly, this figure is closely related to Figs.~\ref{layerMAEfig} and \ref{comparison4}: for small $s$, the Co contributions $\{D^{(Co)}_i\}$ show an increasing tendency with increasing $s$ for $\mid \! i \mid=1$ and $\mid \! i \mid \ge 7$, while they decrease for $2 \le \mid \! i \mid \le 5$. Apparently, this latter effect overcomes the former one, leading to the relatively large value of $K_{Pt}(0)=-1$~meV.\\*

\begin{figure}[htp]
\begin{center}
\includegraphics[scale=0.35]{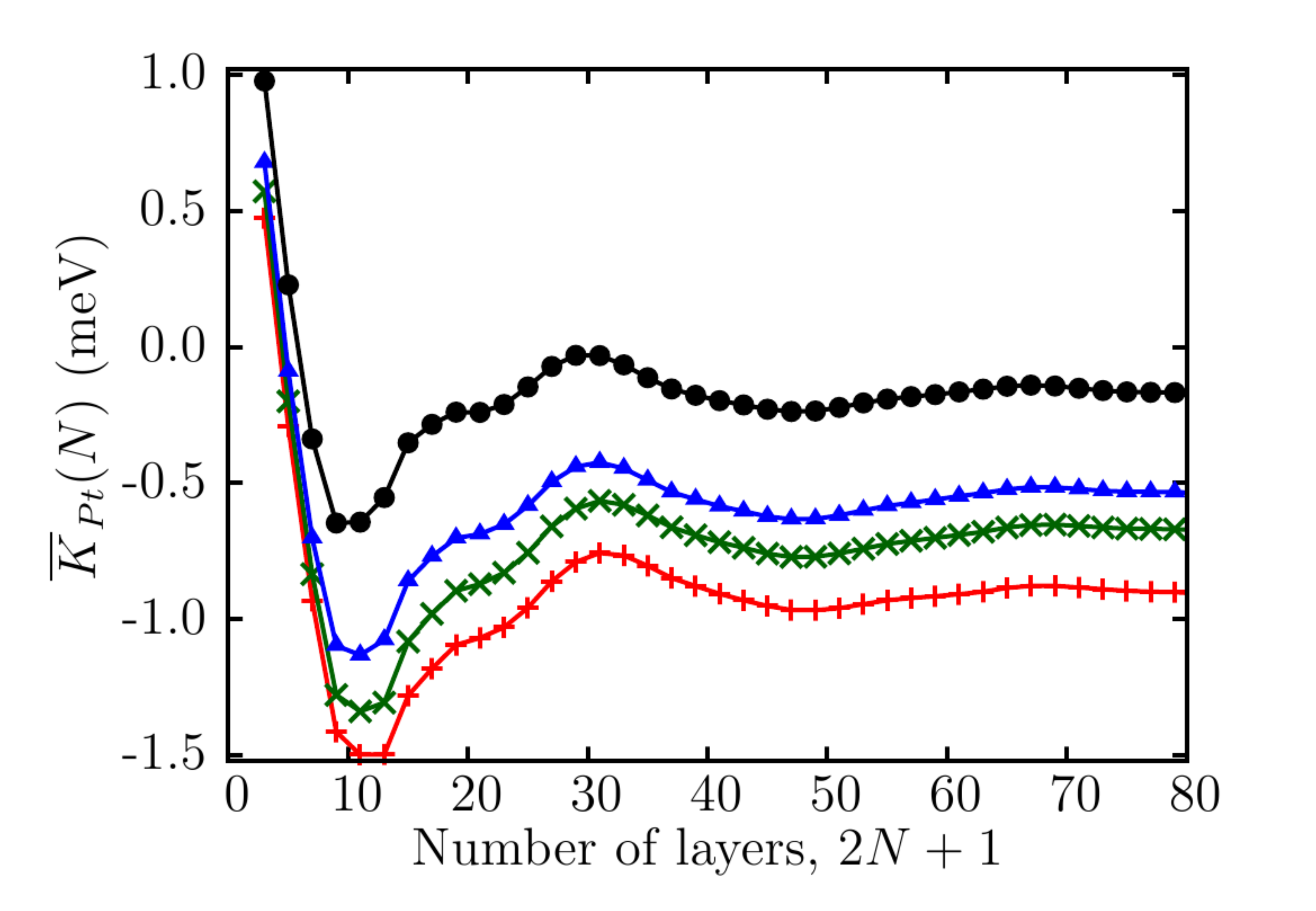}
\caption{(Color online) Calculated excess MAE per platinum atom, $\overline{K}_{Pt}$, as a function of the number of planes $N$
included in the sum in Eq.~(\ref{KPteqsum}) for $s=0.01$ (red $+$), $s=0.05$ (green $\times$), $s=0.10$
(blue $\blacktriangle$) and $s=0.20$ (black {\large $\bullet$}). \label{friedelfig}}
\end{center}
\end{figure}

We also investigated possible effects of the Friedel oscillations on the MAE by truncating the sum in Eq.~(\ref{deltaKeq}) and considering the variation in $\overline{K}_{Pt}$ against the number $N$ of Co planes included in the sum, \begin{equation}
  \overline{K}_{Pt}(s,N) =  \frac{1}{s}\left( sD_{0}^{(Pt)} + (1-s)D_{0}^{(Co)}  + 2\sum_{i=1}^{N}D_{i}^{(Co)} - (2N+1) K_{Co} \right) \; ,
\label{KPteqsum}
\end{equation}
for $1 \leq N\leq 40$. The function $\overline{K}_{Pt}(N)$ for $s=0.01$, $0.05$, $0.10$ and $0.20$ is shown in Fig.~\ref{friedelfig}. For all cases, the maximum of $\overline{K}_{Pt}(N)$ occurs at $2N+1=3$ planes, i.e.~including only one Co layer on each side of the impurity layer. This is because the induced effect on $D_1^{(Co)}$ by the addition of platinum is strongly positive for all $s$, see Figs. \ref{layerMAEfig} and \ref{comparison4}(b).  There is a significant minimum in the calculated $\overline{K}_{Pt}(N)$ at $2N+1\approx 11$ planes. Comparing with Fig.~\ref{layerMAEfig}, it is obvious that this minimum is due to the reduction of $D_i^{(Co)}$ for $3\leq \mid \! i \mid \leq 5$ caused by the addition of Pt. $\overline{K}_{Pt}(N)$ then exhibits a local maximum at $2N+1\approx 31$, mostly due to the Co contributions in layers $8\leq \mid \! i \mid \leq 15$ counterbalancing the Co contributions of opposite sign in layers $3\leq \mid \! i \mid  \leq 5$. Concerning the overall accuracy of the calculated MAE, the effects of the Friedel oscillations remain significant for about $2N+1 < 70$ planes, i.e. for 35 layers on either side of the impurity layer.  In general, the variation in $\overline{K}_{Pt}$ with $N$ spans more than 1.5 meV for all values of $s$.  This of course has significant implications for measuring the MAE in thin film samples, as up to about $2N+1 < 40$ planes, i.e. for film thicknesses $d < 8$~nm, the change in the MAE induced by the Pt impurities located at the centre of the sample, is expected to be extremely sensitive on the film thickness. Note that this finite-size effect is superimposed on and, most likely, amplified by quantum interferences arising from the boundaries of the finite film sample.\\*

It should be noted that, being a mean-field approach, the CPA completely neglects both structural and electronic short-range order effects. Such short-range order effects are likely to be most strongly pronounced for small Pt concentrations $s$. Therefore, our results in the low-$s$ limit should be tested against another method. By employing a fully relativistic real-space embedded cluster Green's function technique as combined with the SKKR method \cite{lazarovits02} we made an attempt to test the effect of electronic relaxations around a Pt impurity placed in hcp bulk Co. We performed self-consistent calculations for a cluster containing a Pt impurity and the neighboring Co atoms up to three nearest neighbor (NN) distances of the hcp lattice ($a_{NN}$). Note that our embedded cluster included 158 Co atoms around the central Pt atom, sorted out geometrically as follows: {\em (i)}  36, 2 $\times$ 30, 2 $\times$ 19 and 2 $\times$ 12 Co atoms in layers 0, $\pm$1, $\pm 2$ and $\pm 3$, and {\em (ii)} 12, 56 and 158 Co atoms within the spheres centered around the Pt site having the radii of $a_{NN}$, $2a_{NN}$ and $3a_{NN}$, respectively. The MAE for this cluster was again calculated using the magnetic force theorem. By summing up all the site-resolved contributions of the MAE in the cluster within the spheres as mentioned above, we obtained the values for $K_{Pt}$, 0.67 meV, -0.31 meV and -0.38 meV, respectively. Obviously, a direct comparison of these values with those for the layered system is hardly possible, since, according to the geometrical classification {\em (i)}, the $K_{Pt}$ for the cluster refers to an incomplete summation over sites in the respective layers. Nevertheless, comparing with the values in Fig.~\ref{friedelfig} related to $2N+1 = 3, 5$ and 7 and for the concentration of $s=0.01$ being closest the case of an impurity, we can conclude that both the trend and the magnitude of $K_{Pt}$ are in satisfactory agreement between the CPA and the real-space calculations.

\section{Summary and Conclusions}
Using the fully relativistic screened Korringa-Kohn-Rostoker method combined with the coherent potential approximation, we have studied the MAE of a bulk hcp Co system in which one of the (0001) atomic plane has been alloyed with Pt  for the whole concentration range, $0 < s \le 1$. We conclude that low concentrations of platinum reduce the overall MAE of this system and that the origin of this reduction are the induced changes in the MAE contributions from the Co atoms. In the limit $s\rightarrow 0$ (bulk Co), the change in MAE per platinum atom added is approximately $-1$ meV.  At larger concentrations, the direct MAE contribution of the platinum, which is positive for all Pt concentrations, starts to increase, but the overall change in MAE due to the addition of platinum is still dominated by induced Co contributions. Interestingly, in the limit of a completely filled Pt layer, addition of one Pt atom increases the MAE of the system by about 2.5 meV. We also investigated the effect of long-ranged Friedel oscillations and established a large sensitivity of the MAE on the number of Co layers included in the calculations. This might have a significant impact on the experimental determination of the MAE in thin film samples of this type of system.

\ack

CJA is grateful to EPSRC and to Seagate Technology for the provision of a research studentship. Financial support was in part provided by the New Sz\'echenyi Plan of Hungary Project ID.~T\'AMOP-4.2.2.B-10/1--2010-0009 and the Hungarian Scientific Research Fund (contracts OTKA PD83353, K77771). KP benefited the Bolyai Research Grant of the Hungarian Academy of Sciences. Support of the HAS Wigner Research Centre of Physics through the usage of its computational facilities is also kindly acknowledged.\\*

\end{document}